\documentclass[aps,prl,twocolumn,showpacs,superscriptaddress]{revtex4-1}
\usepackage[latin1]{inputenc}
\usepackage{graphicx}
\usepackage{epstopdf}
\usepackage{epsfig}
\usepackage{subfigure}
\usepackage{xfrac}
\usepackage{indentfirst}
\usepackage{amsmath, amsthm, amssymb} 
\usepackage{bm}
\usepackage{float}
\usepackage[colorlinks=true, a4paper=true, pdfstartview=FitV, linkcolor=blue, citecolor=blue, urlcolor=blue]{hyperref}
\usepackage{hyperref}
\usepackage{cleveref}
\usepackage{braket}
\begin{document}

\title{Emergence of Spatial Order in Highly Interacting Rydberg Gases}

\author{Jo\~{a}o D. Rodrigues}
\email[Corresponding author:]{joaodmrodrigues@tecnico.ulisboa.pt}
\affiliation{Instituto de Plasmas e Fus\~{a}o Nuclear, Instituto Superior T\'{e}cnico, Universidade de Lisboa, 1049-001 Lisbon, Portugal}
\affiliation{Instituto de F\'{i}sica de S\~{a}o Carlos, Universidade de S\~{a}o Paulo, Caixa Postal 369, 13560-970, S\~{a}o Carlos, S\~{a}o Paulo, Brasil}
\author{Hugo Ter\c cas}
\affiliation{Instituto de Plasmas e Fus\~{a}o Nuclear, Instituto Superior T\'{e}cnico, Universidade de Lisboa, 1049-001 Lisbon, Portugal}
\author{Luis F. Gon\c calves}
\affiliation{Instituto de F\'{i}sica de S\~{a}o Carlos, Universidade de S\~{a}o Paulo, Caixa Postal 369, 13560-970, S\~{a}o Carlos, S\~{a}o Paulo, Brasil}
\author{Luis G. Marcassa}
\affiliation{Instituto de F\'{i}sica de S\~{a}o Carlos, Universidade de S\~{a}o Paulo, Caixa Postal 369, 13560-970, S\~{a}o Carlos, S\~{a}o Paulo, Brasil}
\author{Jos\'{e} T. Mendon\c ca}
\affiliation{Instituto de Plasmas e Fus\~{a}o Nuclear, Instituto Superior T\'{e}cnico, Universidade de Lisboa, 1049-001 Lisbon, Portugal}
\begin{abstract}
\begin{footnotesize}
We describe the emergence of strong spatial correlations, akin to liquid-like behavior and crystallization effects, in low (one and two) dimensional gases of cold Rydberg atoms. The presence of an external electric field permanently polarizes the atoms, which became highly correlated due to the long-range dipole-dipole interaction. We describe a theoretical approach particularly suited for strongly coupled systems and numerically obtain both the two-particle distribution function and the static structure factor. The experimental implementation of such highly interacting systems is discussed, including detailed calculations of the interaction strength for different Rydberg states. The results provide new insights into many-body effects associated with strongly interacting Rydberg atoms, including the possibility of observing novel highly ordered phases.
\end{footnotesize}
\end{abstract}
\maketitle
%
\par
Rydberg atoms \cite{gallagher, Marcassa2014} provide a powerful platform to study strongly interacting systems. One manifestation of such interactive character is the so-called blockade effect, where Rydberg states strongly inhibit further excitation of their neighbors, either due to van der Waals \cite{Tong2004, Singer2004} or dipole-dipole interactions \cite{ Urban2009, Gaetan2009}, which has been exploited, for instance, in the implementation of quantum gates for information processing \cite{Saffman2010}. Moreover, the interactions between the high lying electronic states can be mapped onto light fields, giving rise to strong nonlinear optical effects (photon-photon interactions) \cite{Pritchard2010, Sevincli2011}, even at the single photon level \cite{Peyronel2012}, allowing for the storage and manipulation of quantum optical states in highly excited collective states (Rydberg polaritons) \cite{Maxwell2013}. Moreover, the combination of the strong interactive character with the easy manipulation of Rydberg states constitutes an ideal approach for quantum simulation \cite{Weimer2010, Labuhn2016} and general many-body physics \cite{Heidemann2007, Gurian2012}.
\par
Here, we describe a rather distinct manisfestation of the interactive character of Rydberg atoms, namely the emergence of strong spatial correlations in low-dimensional gases. While many studies have been dedicated to the dynamics and transport of Rydberg excitations \cite{Mourachko1998, Weimer2010, Schausz2012, Gunter2013, Labuhn2016}, in the so-called frozen gas regime, the investigation of the external (motional) degrees of freedom has received less attention. Among the few counter-examples \cite{Wuster2010, Mobius2013, Leonhardt2014}, recent studies \cite{McQuillen2013, Thaicharoen2015, Thaicharoen2016, Goncalves2016, Faoro2016} have demonstrated, however, that Rydberg gases are suitable to the study of the (motional) dynamics associated with highly correlated particle systems, a problem which has been subject to extensive theoretical, computational and experimental research \cite{Ichimaru1982, Bollinger1984, Chu1994, Dubin1999, Knapek2007, Ashwin2010, Ashwin2011, Bannasch2012}. Using integral equation techniques, and surpassing the need to resort to molecular dynamics simulations, we are able to describe the emergence and nature of spatial correlations, in a wide-range of coupling parameters, defined as the ratio of potential to kinetic (thermal) energy, from weakly to strongly correlated regimes. We begin by introducing the integral equation approach for a low dimensional Rydberg gas under dipole-dipole interactions, and obtain both the two-particle distribution function and the structure factor. The results are analyzed in detail and routes towards the experimental obtainment of such highly interacting Rydberg systems are discussed, together with numerical calculations of the coupling parameter for different states.
%
%
%
%
\par
In a permanently polarized sample, Rydberg atoms interact via the general dipole-dipole potential \cite{Jackson, Reinhard2007, Comparat2010, Thaicharoen2016},
\begin{equation}\label{eq:dipole_potential}
V  (r, \theta ) = \frac{C_3}{r^3} \left(1 - 3\cos^2  (\theta ) \right),
\end{equation}
with $\theta$ the angle between the interatomic separation vector $\mathbf{r}$ and the atomic dipoles $\mathbf{P}$ (aligned with external electric field $\mathbf{E}_0$). The strength of the interaction is quantified by $C_3 =  P^2 / 4 \pi \epsilon_0$, with $P$ the atomic dipole moment and $\epsilon_0$ the vacuum electric permitivity. In a one-dimensional sample, the character of the interaction (attractive or repulsive) depends on the polarization angle. In particular, there exists a magic angle $\theta =$ 54.7$^{\circ}$, such that the interatomic potential vanishes \cite{Goncalves2016_2}. In a two-dimensional gas, an isotropic repulsive interaction is obtained when the polarization is perpendicular to the atomic sample. In this case, $\theta = \pi / 2$ and the in-plane potential is simply given by $V(r ) = C_3 / r^3$. The average inter-particle distance in a two-dimensional sample is defined as $a=(\pi n_0)^{-1/2}$, with $n_0 = 1/\pi$ the homogeneous density, in the units of $r/a$. The coupling coefficient may be defined as the mean ratio between potential and kinetic energy, $\Gamma = \langle V(r) \rangle / k_B T$, with $k_B$ the Boltzmann constant and $T$ the temperature of the gas and, in terms of the mean inter-particle distance $\Gamma = C_3 /k_B T a^3$. The weakly and highly correlated regimes, $\Gamma \ll 1$ and $\Gamma \gg 1$, respectively, shall be addressed here.
\par
In general, the (classical) description of strongly coupled systems falls outside the scope of hydrodynamical or Vlasov-like formulations for single-particles densities, where correlations between the atoms are usually ignored. Nevertheless, even in the presence of strong correlations, and for conditions of thermal equilibrium, the full $N$-particle density, $ \rho^{(N)} ( \mathbf{r}^N )$, can be truncated to lower orders, introducing the single and the two particle functions \cite{simple_liquids}, $\rho^{(1)} \left( \mathbf{r}\right) =  \langle \sum_i \delta(\mathbf{r} - \mathbf{r}_i ) \rangle$ and $\rho^{(2)} ( \mathbf{r}, \mathbf{r'} ) = \langle \sum_i \sum_{j \neq i} \delta(\mathbf{r} - \mathbf{r}_i )  \delta(\mathbf{r'} - \mathbf{r}_j ) \rangle$, respectively, where the summations are taken over the total $N$ particles and averaging over the canonical ensemble, relevant for low fluctuations in the number of atoms. We may safely take the single particle density as $\rho^{(1)}(\mathbf{r}) = n_0$ and define the (normalized) two-particle distribution $g^{(2)}(\mathbf{r}_1, \mathbf{r}_2) = \rho^{(2)}(\mathbf{r}_1, \mathbf{r}_2) / n_0^2$. In the case of isotropic interactions $g^{(2)}(\mathbf{r}_1, \mathbf{r}_2) = g^{(2)}( \lvert \mathbf{r}_1 -  \mathbf{r}_2 \rvert) \equiv g(r)$, with the latter usually known as the radial distribution function, such that the average number of particles lying between $r$ and $r+dr$ away from a reference atom is $n_0 g(r) dr$ and $2 \pi r n_0 g(r) dr$ in one and two-dimensional samples, respectively. The emergence of correlation, corresponding to $g(r) \neq 1$, or $h(r) \neq 0$, with $h(r) = g(r) -1$ the pair correlation function, will be fully rooted in the interactions among the Rydberg atoms.
\par
The description of the gas in reciprocal space becomes useful in a number of situations and, in this context, the (static) structure factor is defined as
\begin{equation}\label{eq:structure_factor1}
S(\mathbf{k}) = \frac{1}{N} \langle \rho^{(1)} (\mathbf{k}) \rho^{(1)} (\mathbf{-k}) \rangle = 1 + \frac{1}{N} \langle \sum_{i \neq j} e^{-i \mathbf{k} \cdot \left( \mathbf{r}_i - \mathbf{r}_j \right)} \rangle.
\end{equation}
The last term in Eq. (\ref{eq:structure_factor1}) vanishes when the relative position between any pair of particles, $\mathbf{r}_i - \mathbf{r}_j$, is statistically independent, corresponding to the absence of correlation as $S(\mathbf{k}) = 1$. Moreover, the static structure factor is related with the radial distribution function by
\begin{equation}\label{eq:structure_factor3}
S(k) -1 = n_0 \int d \mathbf{r} e^{-i \mathbf{k} \cdot \mathbf{r}} \left[ g(r) -1 \right],
\end{equation} 
or, equivalently, $S(k) = 1+n_0 h(k)$. It directly describe how materials scatter radiation and, without direct imaging techniques, is usually measured by X-ray or neutron diffraction experiments. It became instrumental in the study of the internal structure of systems like liquid Helium \cite{helium1, helium2, helium3} or strongly correlated plasmas \cite{Ichimaru1982, plasmas2}.
\begin{figure}
\centering
\includegraphics[trim={0 3.8cm 0 1.8cm}, clip, scale=0.21]{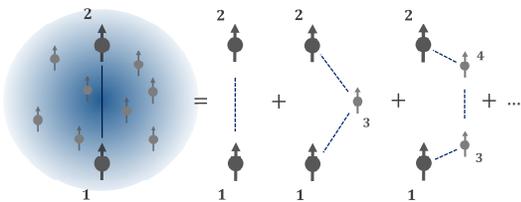}
\caption{(color online) Schematic illustration of the origin of correlation between any pair of dipoles (Rydberg atoms). Besides the direct correlation between 1 and 2, indirect correlations mediated by any number of intermediate particles must be taken into account.}
\label{fig:ornstein_zernike}
\end{figure}
\par
The total correlation between two Rydberg atoms arises both from the direct interaction between them and those with intermediate atoms - see Fig. (\ref{fig:ornstein_zernike}). In this context, we may introduce the direct correlation function $c(r)$, related with the total correlation $h(r)$ by the Ornstein-Zernike relation,
\begin{equation}\label{eq:OZ1}
h(r) = c(r) + n_0 \int d\mathbf{r'} c( \lvert \mathbf{r} - \mathbf{r'} \rvert) h(r').
\end{equation}
While the range of $c(r)$ is usually comparable with that of the pair potential, $V(r)$, the total correlation function is, in general, of higher range, due to the effects of indirect correlations. Although the OZ relation gives an exact relation between the total and direct correlation functions, in order to compute these quantities in a self-consistent manner, some sort of closure condition is needed.
\par
In the absence of drift, pressure and electrostatic forces must balance, $\mathbf{\nabla} \phi =- \frac{\mathbf{\nabla} p}{n}$, with $\phi(r)$ the potential created by a distribution of Rydberg atoms. For an ideal gas at thermal equilibrium $p=n k_B T$ and, $\mathbf{\nabla} \left[ \phi + k_BT \text{ln} (n) \right] = 0$. The solution $n(r) = n_0 \text{exp} \left[ -\phi(r)/k_B T \right]$ is known as the barometric law, which simply determines the distribution around a test particle in the presence of interactions according to a Boltzmann law. Here, an appropriate potential accounting for both the effects of direct and indirect interactions can be obtained by constructing an hierarchy similar to the OZ relation, where the total potential, $\phi(r)$, is the sum of the direct pair-wise term, $V(r)$, and the indirect contribution from any number of intermediate atoms, namely
\begin{equation}\label{eq:hnc_explanation}
\begin{split}
- \frac{\phi(r)}{k_B T} & =  -\frac{V(r)}{k_B T} + n_0 \int d\mathbf{r'} c( \lvert \mathbf{r} - \mathbf{r'} \rvert) h(r') \\
& = - \frac{V(r)}{k_B T} + h(r) - c(r),
\end{split}
\end{equation}
where the last equality follows from the OZ relation. Evoking the barometric law finally yields
\begin{equation}\label{eq:hnc}
g(r) = \text{exp} \left[ - \frac{V(r)}{k_B T} + h(r) - c(r) \right].
\end{equation}
The latter is known as the hyper-netted chain (HNC) closure relation. Together with the OZ relation in Eq. (\ref{eq:OZ1}) forms a closed set of equations for the total and direct correlations functions, $h(r)$ and $c(r)$, respectively. We should stress that, while the OZ relation is exact, the HNC closure constitutes an approximation rooted at the barometric assumption, performing particularly well for systems displaying long-range interactions.
%
%
%
%
\begin{figure}
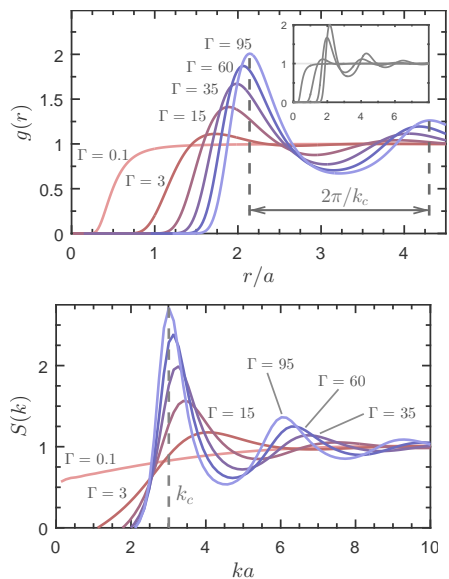

\centering
\begin{subfigure}
\centering
\includegraphics[trim={0 0 0 0}, clip, scale=0.66]{plot_radial_distribution_function_2d}
\end{subfigure}
\vspace{0cm}
\begin{subfigure}
\centering
\includegraphics[trim={0 0 0 0}, clip, scale=0.66]{plot_static_structure_factor_2d}
\end{subfigure}
\caption{(color online) Two-dimensional Rydberg gas. Top-panel - Radial distribution function for different values of the coupling parameter, displaying the emergence of liquid and quasi-periodic correlations, with $k_c$ the wavenumber associated with such regularity in the spatial arrangement. The inset plot depicts the oscillatory character of $g(r)$ and the high range persistence of the spatial correlations. Bottom panel -  Static structure factor for different values of $\Gamma$. The developing peak at approximately $k_c = 2 \pi / \lambda_c \simeq 3a^{-1}$ entails the emergence of stronger correlations and the short-range oscillatory behavior of the radial distribution function.}
\label{fig:2d_results}
\end{figure}
\par
The obtainment of the radial distribution function, $g(r)$, implies a numerical approach to the coupled OZ and HNC closure equations. The former becomes more easily expressed in reciprocal space, $h(k) = c(k) + n_0 c(k) h(k)$. For implementation purposes, the indirect correlation function shall be defined as $e(r) = h(r) - c(r)$. The algorithm begins with an initial guess for the direct correlation function, $c_0(r)$. From its Fourier Transform, $c_0(k)$, we compute the indirect correlation function, $e_0(k)$, using the OZ relation in reciprocal space. Finally, from its inverse Fourier Transform, $e_0(r)$, together with the HNC closure equation, we compute the updated direct correlation function, $c_1(r)$ and repeat the entire procedure until convergence is obtained. In general, such numerical scheme is numerically unstable and, to mitigate this effect, we only mix a small part of the new direct correlation function into the old one. 
\par
The results for a two-dimensional gas are depicted in Fig. (\ref{fig:2d_results}). For weak correlations, $\Gamma \ll 1$, the only prominent feature is the absence of Rydberg atoms close to the reference particle, located at $r=0$, due to the repulsive character of the dipole-dipole interaction. For higher values of the coupling parameter, however, we observe the emergence of strong spacial correlations. Besides the greater region of atom depletion near $r = 0$, due to stronger repulsion, a prominent density peak appears located at approximately twice the mean inter-particle distance, $r \sim 2a$, followed by successive regions of depletion and accumulation of Rydberg atoms, regularly spaced by approximately $\lambda_c \simeq 2 a$. Such oscillatory behavior of $g(r)$ is characteristic of highly correlated systems \cite{Ichimaru1982} and entails the cross-over between liquid and crystalline-like behavior, associated with the high interctive character of the system emerging at higher values of $\Gamma$ \cite{Chu1994, Dubin1999, Ashwin2011}. The emergence of such an highly ordered configuration is also visible in the behavior of the static structure factor, with the growing peaks at multiples of approximately $k_c \simeq 2\pi / \lambda_c$.
\begin{figure}
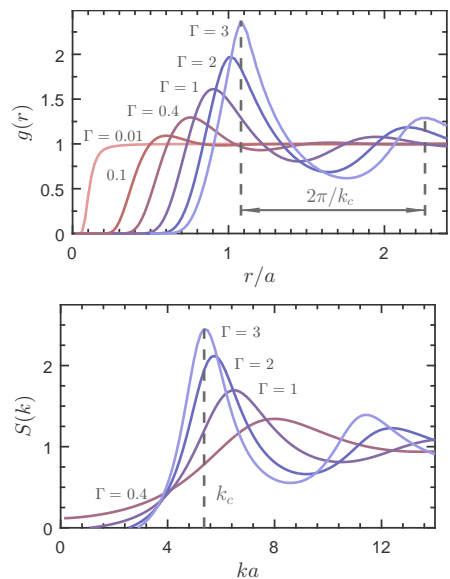

\centering
\begin{subfigure}
\centering
\includegraphics[trim={0 0 0 0}, clip, scale=0.66]{plot_radial_distribution_function_1d}
\end{subfigure}
\vspace{0cm}
\begin{subfigure}
\centering
\includegraphics[trim={0 0 0 0}, clip, scale=0.66]{plot_static_structure_factor_1d}
\end{subfigure}
\caption{(color online) One-dimensional Rydberg gas. The radial distribution function and static structure factor are depicted in the top and bottom panels, respectively, for different values of $\Gamma$. Strong spatial correlations emerge at weaker coupling, due to the reduced dimensionality.}
\label{fig:1d_results}
\end{figure}
\par
In one-dimensional systems, we observe the same kind of spatial order emerging for stronger coupled systems - see Fig. (\ref{fig:1d_results}). Here, however, lower interaction strengths are required to observe similar configurations, due to the reduced degrees of freedom available for spatial arrangement. In other words, the reduced dimensionality makes it harder to create disorder in the system. In fact, no major overhead of interaction induced ordering to thermal induced disordering is required to observe the emergence of strong spacial correlations, which begin to appear for much lower values of $\Gamma$, on the order of unit. A worthy observation is related with the long-range character of the emerging order. Particularly for the one-dimensional case, the height of the first peak in the radial distribution function grows faster than those of higher order. This essentially means that the dipole-dipole interaction are at the origin of short to medium-range order, while true long-range order does not fully arise. This kind of behavior is partially explained by the Mermin-Wagner theorem of statistical physics, which essentially states that for $d\leq2$, with $d$ the dimensionality of the system, low momenta thermal fluctuations destroy true long-range order \cite{Mermin1966, Hohenberg1967}. It is also worth mentioning that, strongly interacting dipolar gases have been previously investigated in the context of polar molecules \cite{Buchler2007}, with quantum Monte Carlo simulations describing the different ground state phases, emerging at high densities and temperature close to 1 $\mu$K. Here, however, and due to the higher temperatures and lower densities inherent to the Rydberg blockade effect, we describe how dipole-dipole interactions are at the origin of strong correlations and many-body dynamics at the classical level.
%
%
%
\par
\begin{figure}
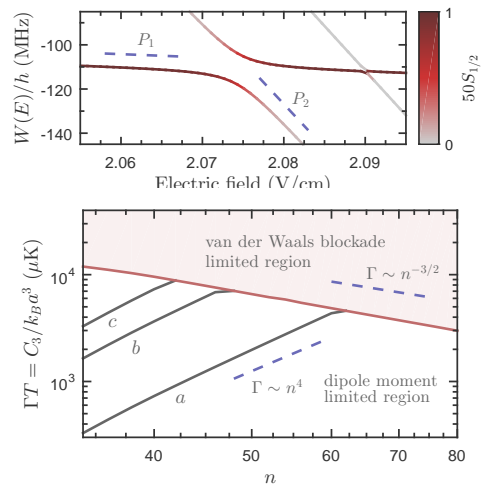

\centering
\begin{subfigure}
\centering
\includegraphics[scale=0.66]{plot_avoided_crossing}
\end{subfigure}
\vspace{0cm}
\begin{subfigure}
\centering
\includegraphics[scale=0.66]{plot_Gamma_function}
\end{subfigure}
\caption{(color online) Top panel - First avoided crossing in the Stark map for the 50$S_{1/2}$ state of $^{85}$Rb, whose energy at zero field defines the axis origin. Here, $P_1$ and $P_2$ correspond of states of low and high dipolar character, respectively and the colored lines indicate the amount of 50$S_{1/2}$ character. Bottom panel - Coupling parameter (normalized to the temperature) as a function of $n$. The a, b and c (black) curves correspond to a ground state density (or Rydberg density, by construction) of $n_g=$ 10$^{9}$, 5x10$^{9}$ and 10$^{10}$ cm$^{-3}$, respectively, where the interaction is limited by the strength of the dipole moment. The red curve depicts the regime limited by the van der Waals blockade, where an on-resonance excitation has been considered, with laser linewidth $\Delta$ = 3 MHz.}
\label{fig:Gamma_function}
\end{figure}

Experimentally, the excitation of Rydberg atoms usually relies on a two-photon transition. For ${^\text{85}}$Rb, the 5$S_{1/2}$ ground state is excited to the intermediate 5$P_{3/2}$ and subsequently promoted to an high-lying Rydberg level n$S_{1/2}$, for instance. An initial low interacting state ensures small interatomic separations, due to blockade effects. For n$S_{1/2}$ levels, the blockade effect only arises from the weaker van der Waals (repulsive) interaction, which becomes significant only for higher $n$ (principal quantum number). Moreover, for $S$ orbitals the van der waals interaction does not depend on the relative atomic orientation, providing fairly homogeneous Rydberg samples. In order to switch into an highly interacting regime, the atoms are transfered to high dipolar states through a Landau-Zener adiabatic passage, promoted by an electric field sweep through an avoided crossing in the Stark landscape \cite{saquet2010, wang2015} - see Fig. (\ref{fig:Gamma_function}). The final highly dipolar linear Stark state becomes permanently polarized in the direction of the electric field, efficiently transferring the atoms from a low to an highly interacting regime \cite{Thaicharoen2016, Goncalves2016}. The permanent dipole moment of a state with energy $W(E)$, with $E$ the electric field, is given by the slope $P = -\frac{dW(E)}{dE}$ and, therefore, large dipole-dipole interactions are achieved by exciting Rydberg levels with large slopes in the Stark map.
\par
While the dipole moment quantifies the strength of the interaction, $C_3 =  P^2 / 4 \pi \epsilon_0$, another equally important parameter for achieving highly coupled samples is the average inter-atomic separation $a$ and, here, two regimes can be distinguished. On the one hand, for high ground state densities and/or high principal quantum number, the van der Waals blockade will limit the Rydberg density. In particular, $V_\text{vdW} = C_6 / r^6$, with the interaction strength smoothly scaling as $C_6 \propto n^{11}$ \cite{Reinhard2007}. In the case of on-resonance excitation, the blockade radius is such that $\Delta = C_6 / r_\text{B}^6$, with $\Delta$ the linewidth of the excitation laser. In this regime, the average inter-atomic distance is then limited by the blockade, $a = r_\text{B}$. For the sake of reference, $r_\text{B}$ = 2.6, 4.2 and 6.0 $\mu$m, for n = 40, 50 and 60, respectively, for the n$S_{1/2}$ states of $^\text{85}$Rb (assuming $\Delta= $3 MHz). On the other hand, for low principal quantum number and ground state densities $n_g$, such that $a=\langle r \rangle = \left( 3 / 4 \pi n_g \right)^{1/3} > r_\text{B}$, with $\langle r \rangle$ the average ground state inter-atomic distance, the van der Waals blockade plays no role and the interaction strength will mostly be limited by the magnitude of the permanent dipole moment.
\par
The coupling parameter $\Gamma$ computed for different principal quantum numbers and normalized to the temperature is depicted in Fig. (\ref{fig:Gamma_function}). Here, the atomic dipole moment is computed for the adiabatic state after the first avoided crossing appearing in the Stark map, obtained by diagonalization of the interaction Hamiltonian. In lowest order, the energy shifts from the linear Stark effect scale as $W \propto E n^2$ and, consequently, the dipole moment $P \propto n^2$ \cite{Reinhard2007}. In the dipole moment limited regime, where $a$ does not scale with the principal quantum number, $\Gamma \propto P^2 \propto n^4$, which is verified by the numerical results. In the van der Waals blockade regime, however, the scaling of the average inter-atomic distance (blockade radius) with $n$ must be taken into account. Since $C_6\propto n^{11}$ and, consequently, $r_\text{B} \propto n^{11/6}$, the coupling coefficient $\Gamma \propto n^{-3/2}$, which is also in agreement with the results. Here, the computation of $C_6$ is performed via second order perturbation theory. Clearly, the fast increase of the blockade radius is more dramatic then the higher dipole moments and, in this regime, the interaction strength decreases with $n$.
\par
Notice that, immediately after the excitation stage, no particular order is expected in the system. This highly out-of-equilibrium initial state shall be followed by a rapid reorganization of the atomic dipoles. This is accompanied by a fast increase in the temperature of the system, as the randomly positioned dipoles are accelerated by the potential landscape into an equilibrium distribution. This mechanism of disorder induced heating has been reported in experiments on ultra cold neutral plasmas (UCNPs) \cite{Killian2007} and could eventually impose a limitation on the achievement of high values of $\Gamma$. In the context of UCNPs, it has been demonstrated that the excitation of initially ordered samples greatly reduces the effects associated with disorder induced heating. This could be obtained either via the Rydberg blockade mechanism \cite{Bannasch2013}, where an initial sample of spatially ordered Rydberg atoms is posteriorly ionized, or via the preordering of atoms in an partially filled optical lattice \cite{Murphy2016}. While the latter could also be employed in the present context, the former more simply means that an initial excitation where the interatomic distance is limited by the blockade mechanism would also suppress the effects associated with disorder induced heating. As recently demonstrated in \cite{Thaicharoen2016}, a Landau-Zener adiabatic crossing is appropriate to create highly interacting samples, with the emergence of the associated spatial correlations and ordered states occurring in just a few microseconds, much faster than the typical lifetimes of the excited states or collisional processes. 
\par
The experimental obtainment of correlation functions involves the detection of the atomic trajectories. Ion-imaging based methods proved to be very efficient in this regard for 3D \cite{Schwarzkopf2011, Schwarzkopf2013} and 1D \cite{Goncalves2016} systems, for both weak van der Waals \cite{Thaicharoen2015} and strong dipole-dipole \cite{Thaicharoen2016} interactions. It allows the retrieval of the spatial distribution of Rydberg atoms, with micrometer resolution and single atom detection capability. The creation of 1D samples can be done by trapping the atoms in a tightly focused dipole trap \cite{Goncalves2016} while, for the 2D case, an astigmatic excitation beam, created by a cylindrical lens, can be used to create a plane of light in the excitation process.
%
%
\par
In summary, we demonstrated the emergence of highly correlated structural phases in low dimensional gases of Rydberg atoms with strong dipole-dipole interactions. The integral equation technique employed here constitutes a general approach to strongly correlated particle systems, particularly well-suited for long-range interactions. Moreover, it surpasses the necessity to resort other commonly used massive numerical approaches, such as Monte Carlo or Molecular Dynamics techniques, together with providing easy manipulation of interaction strengths and overall parameter dependences. In the context of Rydberg gases, it can be applied to different kinds of atom-atom interactions, such as van der Waals or soft-core potentials, the latter corresponding to the case of dressing from far off-resonant excitation, where ground states atoms acquire partial Rydberg character \cite{Pupillo2010, Sevincli2011, Maucher2011, Macri2014}. Moreover, and contrary with other high correlated systems, such as liquids, solid-state materials, plasmas, etc, cold (Rydberg) atom experiments benefit from the high degree of control over parameters such as interacting strength or sample dimensionality. As such, together with the results presented here, we argue that Rydberg atoms provide an ideal platform to the investigation of strongly correlated media and many-body physics, opening a new route towards the simulation of complex systems with cold atom experiments \cite{Bloch2012, Younge2009, Weimer2010, Gurian2012, Labuhn2016}, as well as probing out-of-equilibrium dynamics in strongly interacting systems.
%
%
\par
JDR acknowledges the financial support of FCT - Funda\c{c}\~{a}o da Ci\^{e}ncia e Tecnologia (Portugal) through the grant number SFRH/BD/52323/2013. HT thanks the Security of Quantum Information Group for the hospitality and for partially providing the working conditions, and thanks the support from FCT - Funda\c{c}\~{a}o da Ci\^{e}ncia e Tecnologia (Portugal) through the grant number IF/00433/2015. This work was partially supported by S\~{a}o Paulo Research Foundation (FAPESP) Grants No. 2011/22309-8 and No. 2013/02816-8, the U.S. Army Research Office Grant No. W911NF-15-1-0638 and CNPq.

\bibliographystyle{apsrev4-1}
\bibliography{correlations_rydberg}
\bibliographystyle{apsrev4-1}
\end{document}